\documentclass[aps,prl,reprint,notitlepage,showpacs,nofootinbib,superscriptaddress,twocolumn
,nofloats,preprintnumbers]{revtex4-1}
\usepackage{lineno}
\usepackage{graphicx}
\usepackage[utf8]{inputenc} 
\usepackage{amsmath}
\usepackage{amssymb}
\usepackage{ulem}
\usepackage{float}
\usepackage{comment}
\usepackage{slashed}
\usepackage{subfigure}

\usepackage[T1]{fontenc} 
\usepackage{parskip}
\usepackage{bm}         
\usepackage{array}
\usepackage{multirow}   
\usepackage{booktabs}   

\usepackage{textcomp}
\usepackage{amssymb}
\usepackage{amsmath}
\usepackage{soul}

\usepackage{enumerate}
\usepackage{caption}
\usepackage{ragged2e}
 \DeclareCaptionJustification{justified}{\justifying}
\usepackage{amsmath}
\usepackage{amssymb}
\usepackage{indentfirst}
\usepackage[usenames,dvipsnames]{xcolor}
\usepackage[colorlinks=true,citecolor=darkred,urlcolor=darkred, pdfborder={0 0 0}]{hyperref}
\definecolor{darkred}{rgb}{0.6,0,0}

\definecolor{linkcolor}{rgb}{0,0,0.5}
 
\newcommand {\ignore}[1]{}

\definecolor{bostonuniversityred}{rgb}{0.8, 0.0, 0.0}

\def\gsim{\raise0.3ex\hbox{$\;>$\kern-0.75em\raise-1.1ex\hbox{$\sim\;$}}}
\def\lsim{\raise0.3ex\hbox{$\;<$\kern-0.75em\raise-1.1ex\hbox{$\sim\;$}}}

\usepackage{soul}
\definecolor{mightnightblue}{RGB}{25,25,112}

\definecolor{brown}{rgb}{0.59, 0.29, 0.0}

\newcommand{\eps}{\varepsilon}

\def\21{$\mathrm{SU(2)_L \otimes U(1)_Y}$}

\newcommand{\AddrUNAM}{ {\it Instituto de F\'{\i}sica, Universidad Nacional Aut\'onoma de M\'exico, A.P. 20-364, Ciudad de M\'exico 01000, M\'exico.}}

\newcommand{\AddrLJF}{ {\it Tecnol\'ogico Nacional de M\'exico/ITS de Jerez, C.P. 99863, Zacatecas, M\'exico.}}

\begin{document}

\title{\boldmath\color{BrickRed} New Physics searches in a low threshold scintillating argon bubble chamber measuring coherent elastic neutrino-nucleus scattering in reactors}
\author{E.~Alfonso-Pita}\email{ernestoalfonso@estudiantes.fisica.unam.mx }
\affiliation{\AddrUNAM}
\author{L. J. Flores}\email{ljflores@jerez.tecnm.mx }
\affiliation{\AddrLJF}

\author{Eduardo Peinado}\email{epeinado@fisica.unam.mx}
\affiliation{\AddrUNAM}

\author{E.~V\'azquez-J\'auregui}\email{ericvj@fisica.unam.mx }
\affiliation{\AddrUNAM}

\vspace{0.7cm}


\begin{abstract}
\vspace{0.3cm}
\noindent The sensitivity to New Physics of a low threshold scintillating argon bubble chamber measuring coherent elastic neutrino-nucleus scattering in reactors is reported. Namely, light scalar mediators, sterile neutrino oscillations, unitarity violation, and non-standard interactions are studied. The results indicate that this detector could be able to set stronger constraints than current limits set by the recent COHERENT measurements.
Considering the best scenario, a 100 kg detector located 30 m from a 2000 MW$_{th}$ reactor, a sterile neutrino search would cover most of the space parameter allowed from the reactor anti-neutrino anomaly fit. Unitarity violation studies could set constraints on  $\alpha_{11}$ more stringent than the current oscillation experiments fit. 
A low threshold argon detector with very low backgrounds has the potential to explore New Physics in different scenarios and set competitive constraints.
\end{abstract}

\maketitle

\section{Introduction}

\noindent
Coherent Elastic neutrino-Nucleus Scattering (CE$\nu$NS) is a Standard Model (SM) process that has attracted interest from the Physics community. CE$\nu$NS offers the possibility to perform high precision measurements in several processes of the SM \cite{Fernandez-Moroni:2020yyl, Coloma:2020nhf, Miranda:2020tif, Cadeddu:2021ijh} as well as exploring New Physics (NP) scenarios in nuclear and particle physics \cite{Mustamin:2021mtq, Miranda:2021kre, AristizabalSierra:2021kht, Dasgupta:2021fpn, Colaresi:2021kus}. CE$\nu$NS was first observed by the COHERENT collaboration using a CsI[Na] crystal \cite{COHERENT:2017ipa} and later with a LAr detector \cite{COHERENT:2020iec} in the Spallation source SNS at Oak Ridge. This elusive process is still pending to be observed for neutrinos produced in nuclear reactors and it is currently a race among several collaborations \cite{CONNIE:2019swq, CONUS:2020skt, Belov:2015ufh, Agnolet:2016zir,Strauss:2017cuu, Billard:2018jnl, Akimov:2017hee, Fernandez-Moroni:2020yyl, SoLid:2020cen}. Recently, evidence of its observation has been reported using a germanium detector~\cite{Dresden:2022jc}, indicating that the first measurement of this process for reactor neutrinos is expected within less than a couple of years. The observation of this process requires a device capable to detect low-energy nuclear recoils and achieve low backgrounds operating at low energy thresholds (sub keV) for long time periods.

Several detectors are currently taking data \cite{CONNIE:2019swq, CONUS:2020skt} or under construction \cite{Belov:2015ufh, Agnolet:2016zir,Strauss:2017cuu, Billard:2018jnl, Akimov:2017hee, Fernandez-Moroni:2020yyl, SoLid:2020cen}. The SBC collaboration is developing a low threshold argon scintillating bubble chamber aiming to achieve a 100 eV threshold. This device is insensitive to electromagnetic interactions, which greatly suppresses the majority of the backgrounds~\cite{Giampa_2020,PhysRevD.103.L091301}.

A study of the physics reach of the Scintillating Bubble Chamber detector showed its high sensitivity to the weak mixing angle, neutrino magnetic moment, and the search for a light Z$^\prime$ gauge boson mediator~\cite{PhysRevD.103.L091301}. This work extends the physics potential of this detector to other New Physics scenarios, such as searches for light scalar mediators, sterile neutrinos, unitarity violation, and non-standard interactions. The analysis reported in this manuscript also applies to any other technique reaching 100-eV nuclear recoils, eliminating electron-recoil backgrounds, and scaling to 10–100-kg target masses.

A brief description of the experimental scenarios considered is given in the next section (Experiment Description). The following section (New Physics) describes the analysis methods to extract the sensitivity to NP.
 
\section{Experiment description}

The Scintillating Bubble Chamber is a super-heated detector based on 10 kg of liquid argon (LAr) \cite{Giampa_2020, PhysRevD.103.L091301, PhysRevD.103.L091301}. The detector consists of two fused silica vessels, an inner and an outer jar. The target fluid is contained between the jars and this system is immersed in a pressure vessel filled with liquid CF${_4}$ acting as thermal bath and hydraulic fluid. The chamber is designed to operate with a low threshold of 100 eV. This detector has a system of 32 SiPMs (Hamamatsu VUV4 Quads), these sensors record the light created in the LAr and allow background discrimination. In addition, an array of 8 piezoelectric sensors are coupled to the quartz jar to register the acoustic signal created during the bubble formation. Three cameras and lenses are located outside the pressure vessel for imaging of the bubbles produced in the LAr.

A bubble chamber has excellent characteristics for the detection of neutrinos via coherent elastic scattering with argon nuclei. This detector presents high discrimination levels for electromagnetic backgrounds and sub-keV operating threshold allowing the detection of nuclear recoils induced by CE$\nu$NS. 
 
Three experimental setups are explored in this work, assuming one year of livetime. Setup A assumes a 10-kg LAr chamber located 3 m from a 1 MW$_{th}$ reactor and setup B considers a 100-kg LAr chamber located at 30 meters from a 2000 MW$_{th}$ power reactor. Setups A and B use 2.4\% uncertainty in the anti-neutrino flux. Setup B(1.5) is the same as setup B but uses a 1.5\% uncertainty in the anti-neutrino flux. Details of the experimental setups are presented in \cite{PhysRevD.103.L091301}. Calibration strategy for the detector and backgrounds estimated with a GEANT4 \cite{Geant4_0, Geant4_1, Geant4_2} Monte Carlo simulation used for the study described in this manuscript are described elsewhere \cite{PhysRevD.103.L091301}. 
 
A TRIGA Mark III research reactor located at the National Institute for Nuclear Research (ININ) near Mexico City is being explored as a possible location for setup A. This reactor is movable, located inside a water pool, that would allow baselines between 3 and 10 m. The Laguna Verde (LV) power reactor consisting of two BRW-5 (Boiling Water Reactors) units located on the east coast of Mexico in the Gulf of Mexico is also explored as a possible location for setups B.
  
\section{New Physics}

The potential to probe NP scenarios with the 10 kg and 100 kg LAr bubble chambers described in \cite{PhysRevD.103.L091301} is investigated. A 100 eV threshold was assumed for these studies as well as a smearing in the recoil energy with a normalized Gaussian function with standard deviation $\sigma = T$, considering a 100\% uncertainty in the measured recoil energy.

The Standard Model cross section for CE$\nu$NS is 
\begin{equation}
\frac{d\sigma}{dT} = \frac{G_F^2}{2\pi}M_N Q_w^2 \left(2 - \frac{M_N T}{E_\nu^2}\right),
\label{eq:crossSec}
\end{equation}
where
\begin{equation}
Q_w = Z g_p^V F_Z(q^2) + N g_n^V F_N(q^2),
\end{equation}
and $m_N$, $Z$, $N$ are the nuclear mass, proton, and neutron number of the detector material. The number of events is calculated by convoluting the cross-section with the reactor anti-neutrino spectra. The theoretical prediction of the Huber$+$Mueller model~\cite{Huber:2011wv,Mueller:2011nm}, with a $2.4\%$ uncertainty, is considered. The uncertainties in the form factors are negligible with respect to the uncertainty in the anti-neutrino spectra~\cite{Tomalak:2020zfh}.

A fit with the following $\chi^2$ function is performed
\begin{equation}\begin{array}{lcl}
\chi^2 &=&\underset{\alpha,\beta}{\min}\left[\left(\frac{N_\mathrm{meas} - (1+\alpha)N_\mathrm{th}(X)- (1+\beta)B}{\sigma_\mathrm{stat}} \right)^2\right.\\ &&\left. + \left(\frac{\alpha}{\sigma_\alpha}\right)^2 + \left(\frac{\beta}{\sigma_\beta}\right)^2\right],\label{eq:chisq}\end{array}\end{equation} 
where $N_\mathrm{meas}$ is the measured events, $N_\mathrm{th}(X)$ is the theoretical prediction, $B$ is the background coming from the reactor,  $\sigma_\mathrm{stat}$ is the statistical uncertainty, and $\sigma_{\alpha, \beta}$ are the systematic uncertainties on the signal and background, respectively. The variable $X$ refers to the parameter to be fitted. The statistical uncertainty can be expressed as
\begin{equation}
    \sigma_\mathrm{stat} = \sqrt{N_\mathrm{meas} + B_{cosm}},
\end{equation}
where $B_{cosm}$ is the background from cosmogenic neutrons.

The $\chi^2$ function is minimized over the nuisance parameters $\alpha$ and $\beta$. $\alpha$ is used to take into account the systematic errors coming from the anti-neutrino flux. The uncertainty of $2.4\%$ from the Huber$+$Mueller model is translated to $\sigma_\alpha=0.024$. On the other hand, $\beta$ will consider the systematic error due to reactor backgrounds, which for the setup at ININ is taken as $10\%$, namely $\sigma_\beta=0.10$. This parameter is not included for setups B and B(1.5) since the location where the chamber would be placed at LV (30 m from the reactor core) is outside of the reactor building. Backgrounds and systematic uncertainties for the three configurations are thoroughly described in \cite{PhysRevD.103.L091301}.

The following sections describe the New Physics reach of the detector to a light scalar mediator, a sterile neutrino search, unitarity violation, and non-standard interactions.

\section{Light Scalar Mediator}

\begin{figure}[t]
    \centering
    \includegraphics[width=\linewidth]{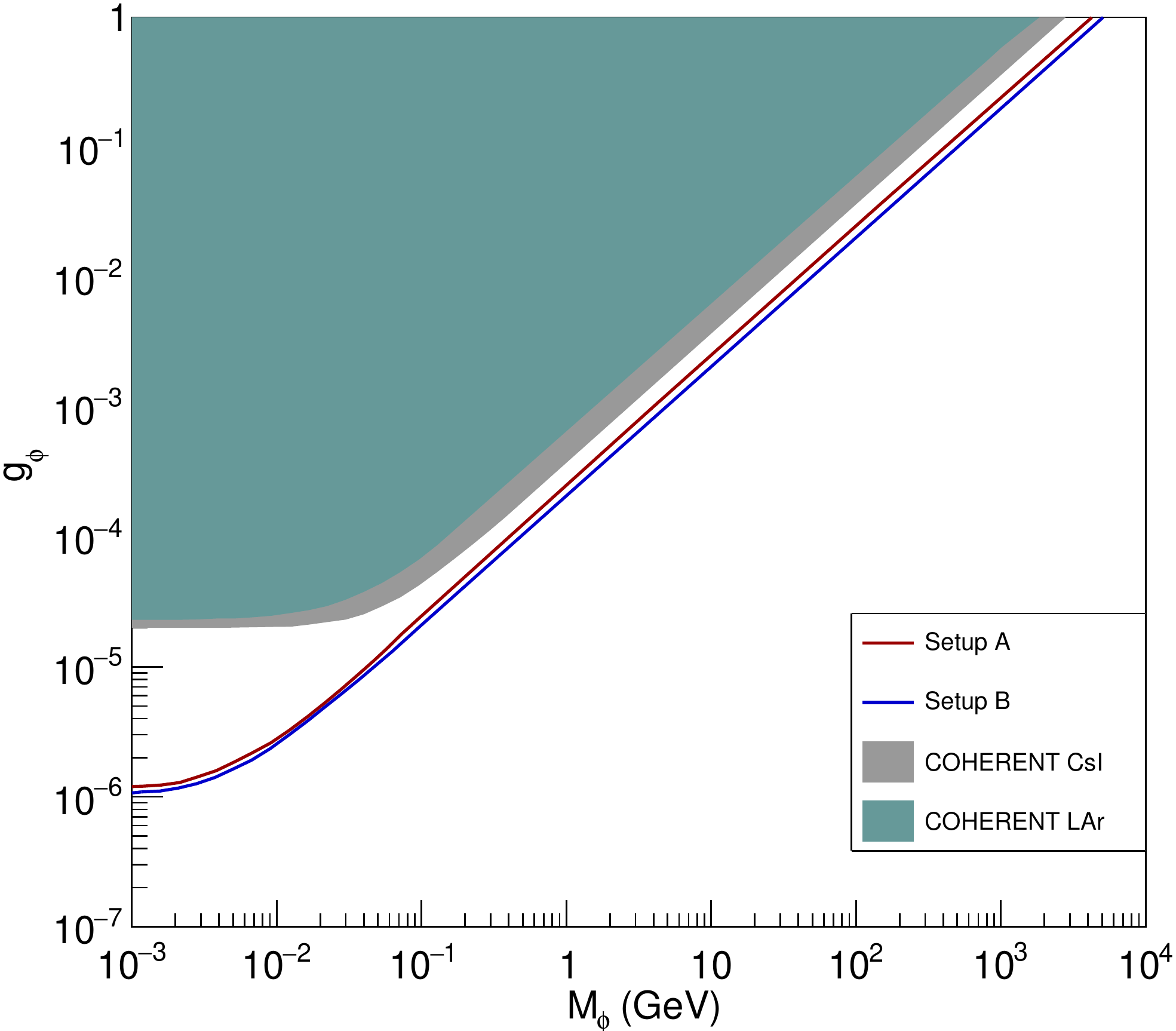}
    \caption{Exclusion regions at $95\%$ C.L. in the $g_\phi-M_\phi$ plane. The solid purple line represents the limit for setup A, while the orange line is the limit for setup B. The shaded brown and yellow regions correspond to the exclusions set by the COHERENT collaboration, using CsI~\cite{Akimov:2017ade} and LAr~\cite{Akimov:2020pdx,Akimov:2020czh} detectors, respectively.}
    \label{fig:scalarMediator}
\end{figure}

A light scalar mediator with universal coupling to quarks and leptons is considered. In such a case, the effective dimension six operator is proportional to the square of the ratio of the coupling over the mass.
The scalar contribution to the CE$\nu$NS cross section is expressed as
\begin{equation}
    \frac{d\sigma}{dT} = \frac{M_N^2}{4\pi}\frac{g_\phi^4 Q_\phi^2 T}{E_\nu^2 (2M_N T + M_\phi^2)^2},
\end{equation}
which is added to the SM contribution without interference. Here, the coupling $Q_\phi$ is a function of the hadronic form factors $f_{T_q}^{p,n}$~\cite{Hoferichter:2015dsa}.
The $95\%$ exclusion regions for setups A and B in the plane $g_\phi-M_\phi$ are shown in Fig.~\ref{fig:scalarMediator}, together with the current limits set by the COHERENT-CsI and -LAr measurements. The setups A and B will give stronger constraints than the current COHERENT data due to the higher number of events expected.

\section{Sterile Neutrino Search}

The three neutrino oscillations have been confirmed by many experiments \cite{Super-Kamiokande:1998kpq, SNO:2002tuh, KamLAND:2002uet} in agreement with the SM with three massive neutrinos. Despite this successful description, there are several experimental results~\cite{LSND:1996ubh, SAGE:2009eeu, MiniBooNE:2013uba} that could extend the current three flavor model, pointing to the existence of at least one additional sterile neutrino. These extra neutrinos are known as "sterile" since they might interact only through the mixing with the active states.
CE$\nu$NS offers a window to search for sterile neutrinos allowing to set constraints in different scenarios. A $3+1$ neutrino hypothesis is explored in this work and sensitivity limits are established by measuring CE$\nu$NS with the scintillating bubble chamber near a reactor for the three setups under study. The $\bar{\nu}_e$ survival probability can be expressed as a function of the propagation length (L), the neutrino energy (E), and the $4\times4$ neutrino mixing matrix as follows: 
\begin{equation}
\begin{split}
    P_{\bar{\nu}_e \rightarrow{} \bar{\nu}_e} \left(\frac{L}{E}\right) = 1 &- 4\sum_{k>j}|U_{ek}|^2|U_{ej}|^2\sin^2\left(\frac{\Delta m^2_{kj}L}{4E}\right),\\
                        = 1 &- \sin^2 2\theta_{13}\sin^2 \Delta_{13}\\
                        &- \sin^2 2\theta_{14}\sin^2 \Delta_{41},
    \end{split}
\end{equation}
where $U_{\alpha i}$ is the neutrino mixing matrix element for flavor $\nu_\alpha$ ($\alpha = e, \mu, \tau$) and mass eigenstate $\nu_i$ ($i = 1, 2, 3$). The neutrino squared-mass differences are represented as $\Delta m_{ij}^2 = m^2_i - m^2_j$ and $\Delta_{ij}$ is a function of the ratio L/E, expressed as follows
\begin{equation}
\Delta_{ij} = 1.267 \Delta m^2_{ij} \left(\frac{L}{E}\right).
\end{equation}

Fig.~\ref{fig:sterile_neutrino_search} shows the limits established by individual analysis for the three setups, and an analysis using the far/near ratio for setup A. This is possible due to the fact that the reactor considered for setup A is movable, inside a water pool. This results in baselines from 3 to 10 m. The far location considered is at 7 m, with the near at 3 m.
\begin{figure}[htpb!]
    \centering
    \includegraphics[width=\linewidth]{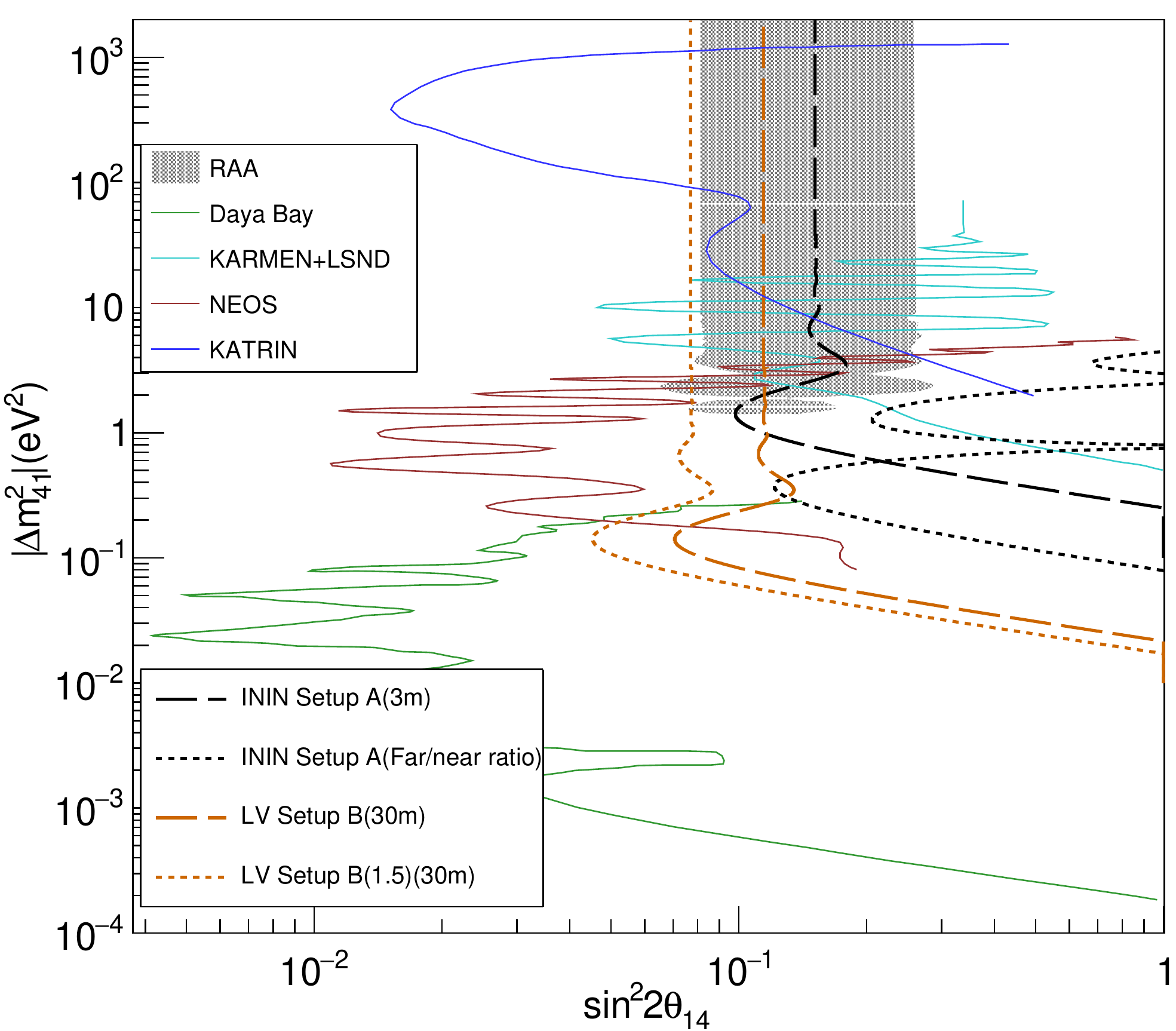}
    \caption{Expected 95$\%$ C.L. exclusion region from sterile neutrino searches. The black dashed and dotted lines represent an expected limit for setup A with a distance of 3 m and for a far/near ratio, respectively. The far(near) location is at 7(3) m. The orange dashed and dotted lines characterize an exclusion sensitivity for setups B and B(1.5), respectively. A comparison of the sensitivity limits is performed with other experiments such as Daya Bay \cite{DayaBay:2016qvc} 95$\%$ C.L. (green), KARMEN+LSND \cite{Conrad:2011ce} 95$\%$ C.L. (light blue), NEOS \cite{NEOS:2016wee} 90$\%$ C.L. (red) and KATRIN \cite{KATRIN:2020dpx} 95$\%$ C.L (dark blue). The shaded region is allowed by the reactor anti-neutrino anomaly (RAA) fit \cite{NEOS:2016wee}, enclosing favored solutions. The rest of the contours disfavor solutions to their right.}
    \label{fig:sterile_neutrino_search}
\end{figure}

Setup B(1.5) will cover all the reactor anomaly and it will give stronger constraints than any other experiment for $10 < |\Delta m^2_{41}| < 80$ eV$^2$. KARMEN+LSND \cite{Conrad:2011ce} alternate the best limits with setup B(1.5) in the region $3 < |\Delta m^2_{41}| < 10$ eV$^2$. These results reflect the importance of reducing the uncertainty in the anti-neutrino flux since that is the only difference between setups B and B(1.5).


\section{Unitarity Violation}

Measuring CE$\nu$NS in reactor allows to study unitarity violation (UV) in the neutrino mixing matrix, predicted by many New Physics scenarios \cite{Miranda:2020syh, Soumya:2021dmy, Forero:2021azc, Wang:2021rsi}, including heavy sterile neutrinos. In this scenario, constraints are set in the non-unitarity parameters through the neutral current. The addition of extra heavy fermions implies the non-unitarity of the $3 \times 3$ light neutrino mixing matrix. In this case, the generalized charged current weak interaction mixing matrix is expressed as
\begin{equation}
    N = N^{UV}\cdot U^{3 \times 3},
\end{equation}
where $N^{UV}$ represents the UV effects corresponding to New Physics, and $U^{3 \times 3}$ is the standard $3 \times 3$ unitary mixing matrix \cite{Rodejohann:2011vc}. The $N^{UV}$ matrix can be parametrized as follows
\begin{equation}
    N^{UV} = 
    \begin{pmatrix}
    \alpha_{11} & 0 & 0\\
    \alpha_{21} & \alpha_{22} & 0\\
    \alpha_{31} & \alpha_{32} & \alpha_{33}
\end{pmatrix},
\label{matrix_UV}
\end{equation}
where the diagonal elements are real numbers and the off-diagonal are complex. For the case of short-baseline experiments, such as the three setups considered in this work, the UV contribution arises from the zero-distance effect. Hence, the survival and transition probabilities of interest considering an electron anti-neutrino source can be express as 
\begin{equation}
\begin{split}
    &P_{ee} = \alpha_{11}^4, \\
    &P_{e\mu} = \alpha_{11}^2|\alpha_{21}|^2,  \\
    &P_{e\tau} = \alpha_{11}^2|\alpha_{31}|^2. \\
 \end{split}   
\label{Oscilatoin_Prob}
\end{equation}
The parameters $\alpha_{11}$, $\alpha_{21}$, and $\alpha_{31}$ are estimated with a $\chi^2$ function identifying the optimal values of the parameters for setups A, B, and B(1.5). Since CE$\nu$NS is flavor blind, namely all the neutrino flavors are detected, $N_\mathrm{th}$ is given by the sum of the UV contributions of the three flavors as
\begin{equation}
N_\mathrm{th} = \alpha_{11}^2(\alpha_{11}^2 + |\alpha_{21}|^2 + |\alpha_{31}|^2)N_\mathrm{meas},
\end{equation} 
where the fit for the $\chi^2$ function is defined in Equation~\eqref{eq:chisq}.

\begin{figure}[htpb!]
    \centering
    \includegraphics[width=\linewidth]{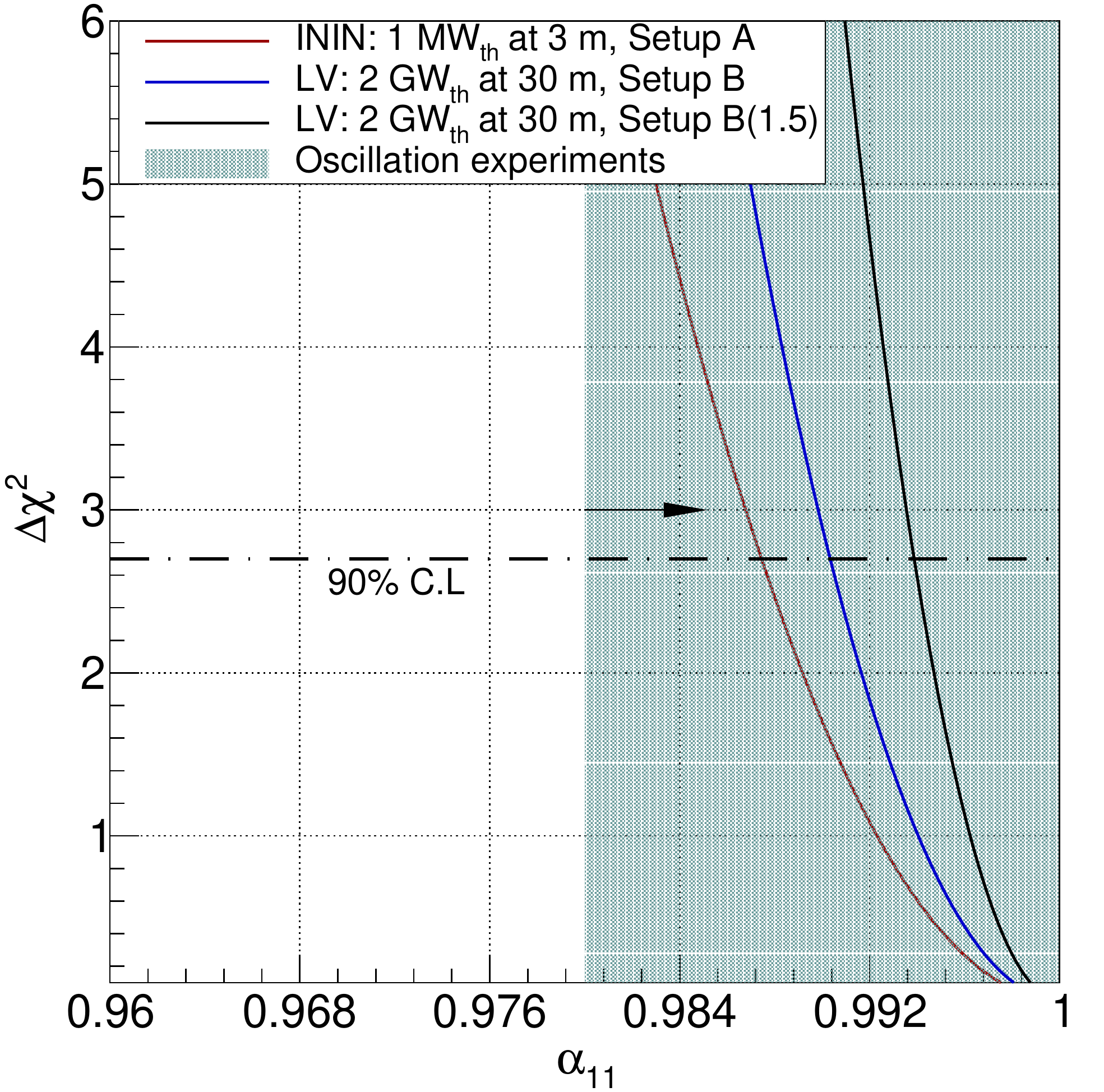}
    \caption{Sensitivity of the diagonal parameter $\alpha_{11}$ for setups A (red), B (blue), and B(1.5) (black). The sensitivity from oscillation data \cite{Miranda:2020syh} is also presented.}
    \label{fig:Alpha_11}
\end{figure}

\begin{figure}[htpb!]
    \centering
    \includegraphics[width=\linewidth]{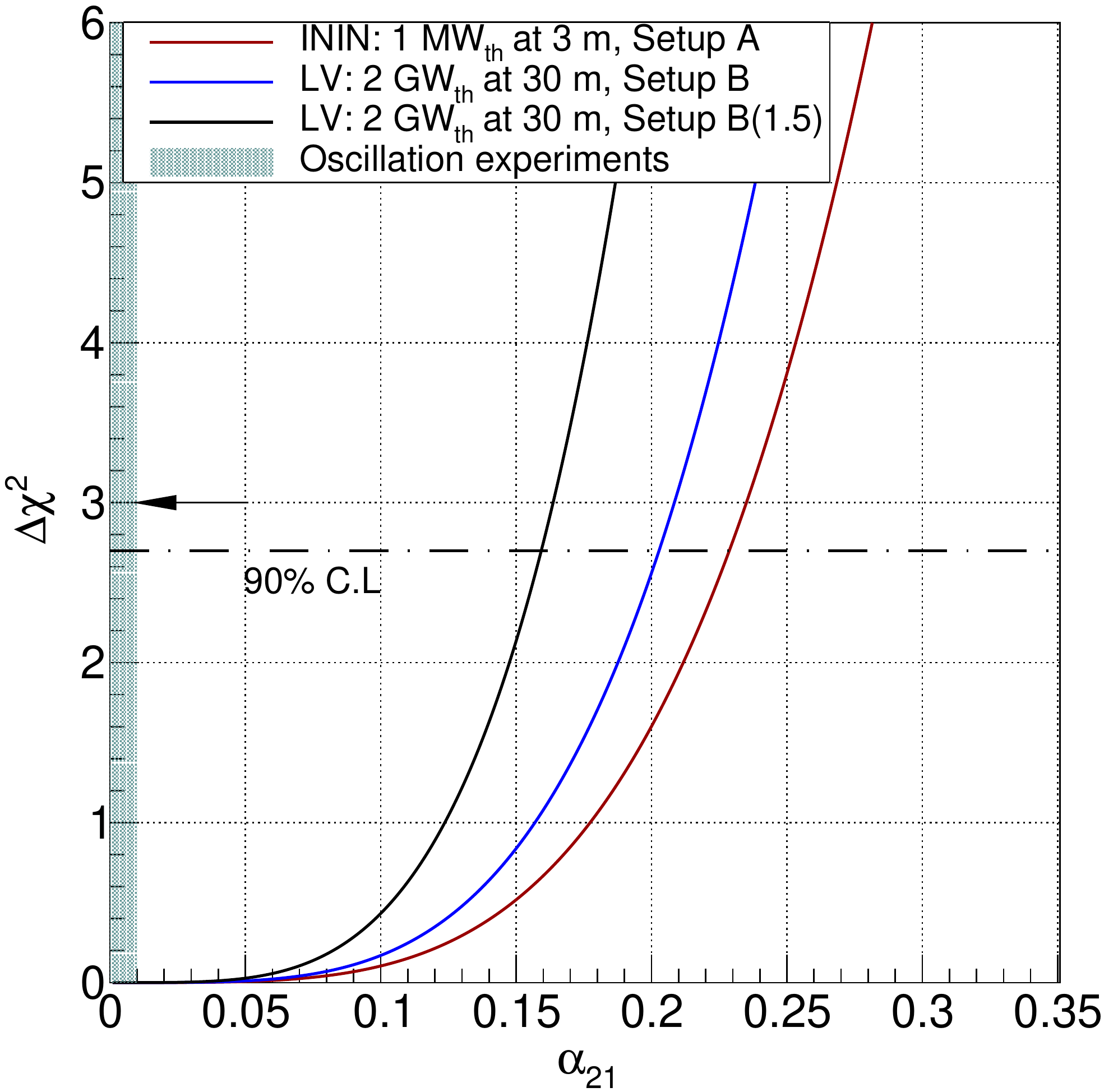}
    \caption{Sensitivity of the non-diagonal parameter $\alpha_{21}$. The projections for setups A (red), B (blue), and B(1.5) (black) are compared with the sensitivity from oscillation data \cite{Miranda:2020syh}.}
    \label{fig:Alpha_21}
\end{figure}

Figs.~\ref{fig:Alpha_11} and~\ref{fig:Alpha_21} show the expected sensitivities of the considered setups, to the diagonal and non-diagonal parameters, respectively, along with the current limits set by global neutrino oscillation data fits \cite{Escrihuela:2016ube}. In addition, Fig.~\ref{fig:sterile_neutrino_search} presents the expected constraints (90$\%$ C.L.) in the $|\alpha_{21}|^2 - |\alpha_{11}|^2$ parameter space. It can be noted that the scintillating bubble chamber could establish stronger constraints than oscillation experiments for the case of $\alpha_{11}$.

\begin{figure}[htpb!]
    \centering
    \includegraphics[width=\linewidth]{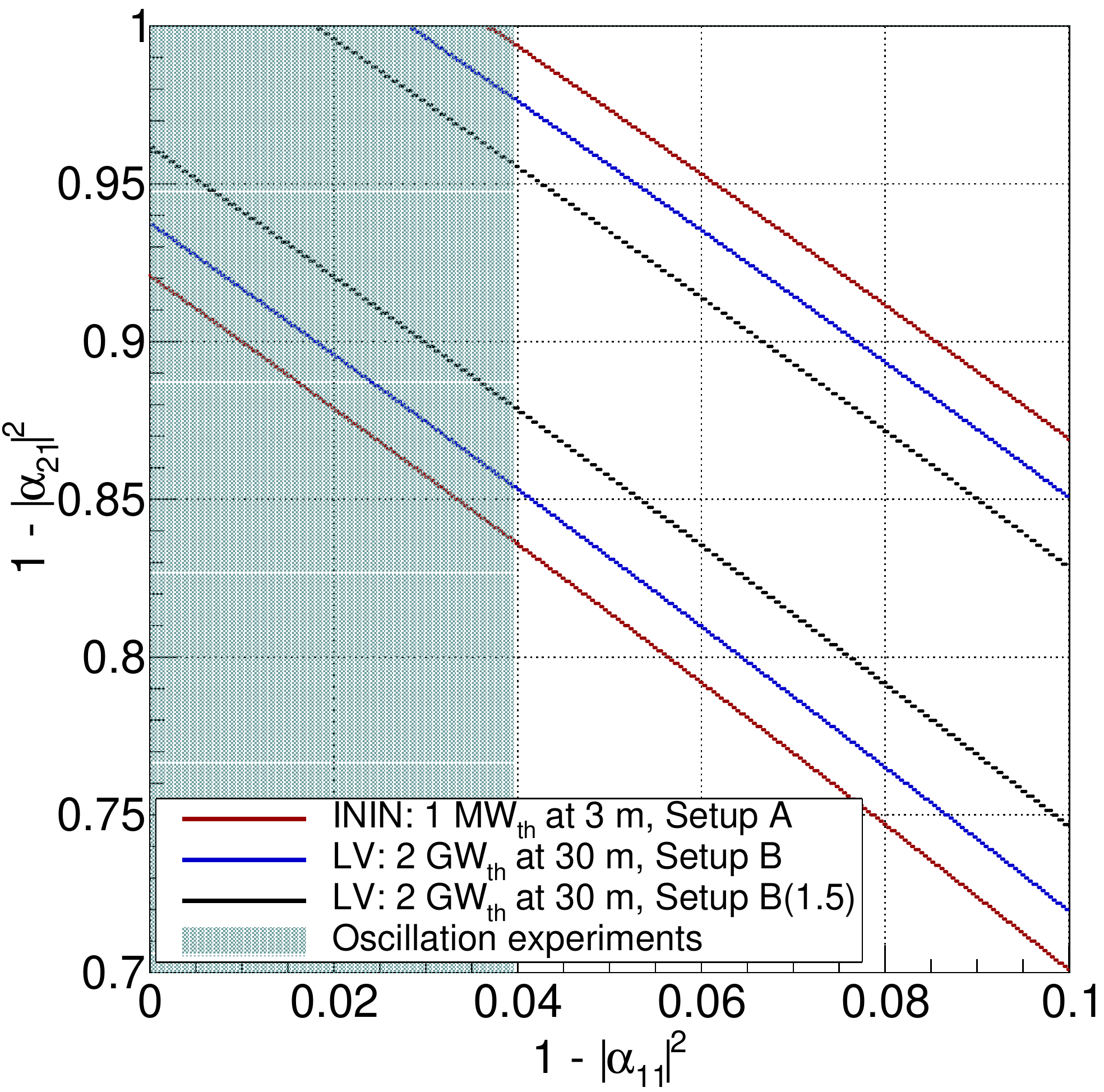}
    \caption{Constraints at 90$\%$ C.L. on the deviations from unitarity in the $|\alpha_{21}|^2 - |\alpha_{11}|^2$ parameter space. Red, blue, and black lines correspond to setups A, B, and B(1.5), respectively. The sensitivity from oscillation experiments \cite{Miranda:2020syh} is also presented.}
    \label{fig:sterile_neutrino_search}
\end{figure}

\section{Non-standard Interactions}

\begin{figure}[htpb!]
    \centering
    \includegraphics[width=\linewidth]{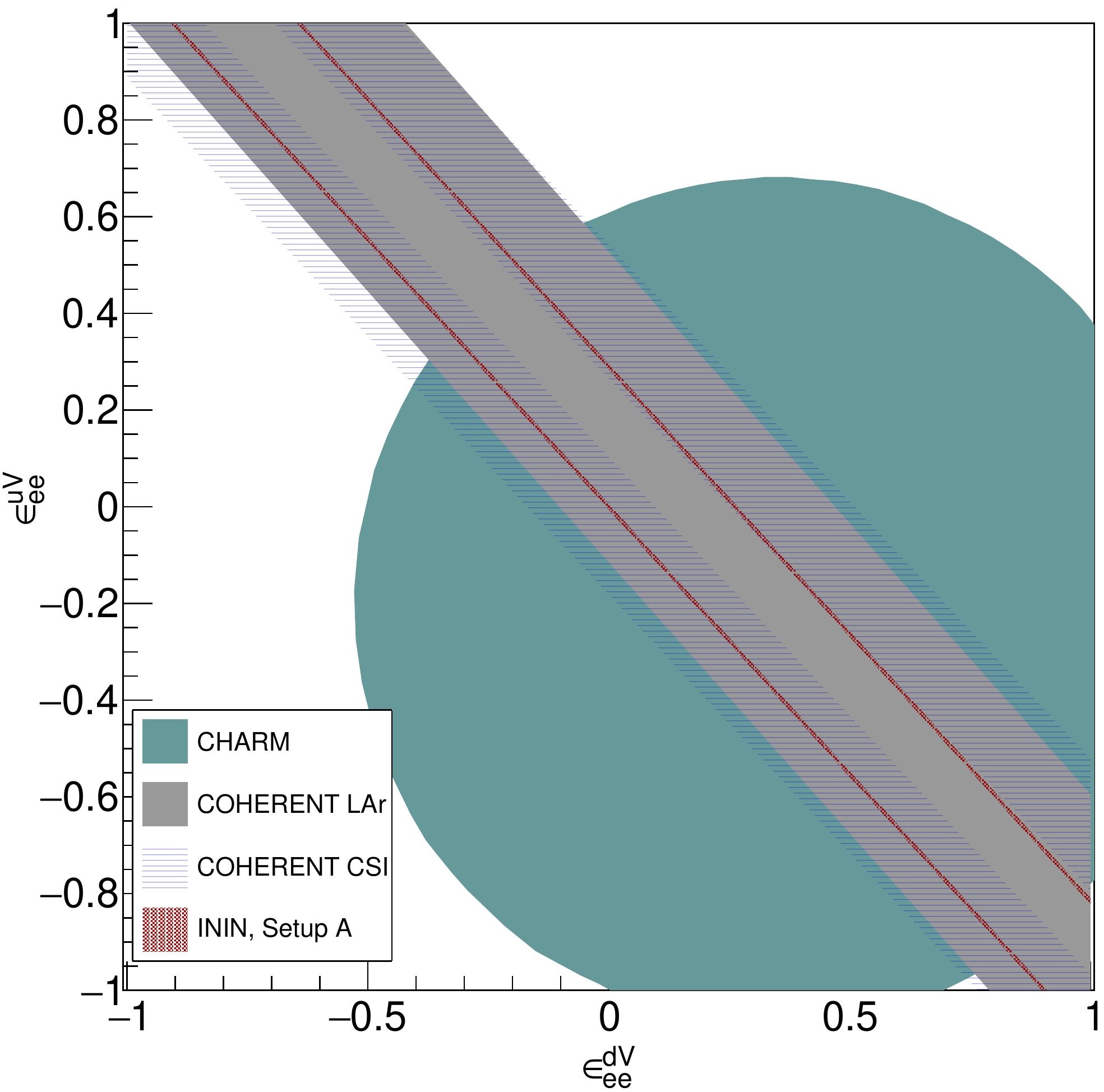}
    \caption{Predicted sensitivity (90$\%$ C.L.) for setup A, represented as two red lines (very narrow bands), for parameters $\eps_{ee}^{fV}$ due to non-standard interactions. Limits established by setups B and B(1.5) cover nearly identical space parameter and are not shown. Allowed parameters established by the CHARM experiment (green) \cite{CHARM:1986vuz} and by the COHERENT experiment using LAr (blue) and CsI (grey) detectors~\cite{COHERENT:2019iyj} are also shown.}
    \label{fig:NSI}
\end{figure}

Any deviation from the SM CE$\nu$NS cross section would hint for physics beyond the Standard Model. The most common modification of the SM Lagrangian is through the non-standard interactions (NSI) formalism, that consists in modifying the neutral current component with the extra contribution:
\begin{equation}
    \mathcal{L}_\mathrm{NC}^\mathrm{NSI} = -2\sqrt{2}G_F \sum_{f,P,\alpha,\beta}  \eps_{\alpha\beta}^{fP}( \bar{\nu}_\alpha \gamma^\mu P_L \nu_\beta)   (  \bar{f}\gamma_\mu P_X f),
\end{equation}
where $f$ represents the $u$ and $d$ quarks, $\alpha$ and $\beta$ correspond to the neutrino flavors $(e, \mu, \tau)$, $P_X$ is the right and left chirality projectors, and $\eps_{\alpha\beta}^{fP}$ represents couplings that characterize the strength of the NSI. Fig. \ref{fig:NSI} presents constrains (90$\%$ C.L.) in the values of the parameters $\eps_{ee}^{fV}$ from the projections of setups A, B, and B(1.5).
The constraints from COHERENT-LAr are the single gray band, while the COHERENT-CsI limits result in the two light-blue stripes bands. Setup A is shown, which is the less restrictive setup for the scintillating bubble chamber. For this scenario, the constraints are no longer one band but two very narrow bands, nearly lines, shown in red color.

\section{Conclusions}

The sensitivity for different New Physics scenarios of a low threshold LAr scintillating bubble chamber measuring CE$\nu$NS in a reactor has been investigated. The work reported in this manuscript shows better sensitivity than the current results established by the COHERENT collaboration, demonstrating the high potential of the bubble chamber technology. Searches for a light scalar mediator can achieve coupling values as low as ${\sim}10^{-6}$ for masses of ${\sim}10^{-3}$ GeV. By combining the setups described in the manuscript, namely a 10 kg detector at 3 m from a 1 MW$_{th}$ reactor and a 100 kg detector at 30 m from a 2000 MW$_{th}$ power reactor, a search for sterile neutrinos in the $|\Delta m^2_{41}|$ range of $10^{-2}-10^3$ eV$^2$ would exclude the majority of the parameter space allowed by the reactor anti-neutrino anomaly. In addition, strong limits on the non-unitarity of the $3\times3$ neutrino mixing matrix are achieved after one year of exposure, competitive with current neutrino oscillation data fits. Lastly, results on non-standard interactions by modifying the neutral current component with the addition of new couplings show complementarity with the results from the CHARM experiment. CE$\nu$NS experiments, and in particular, the scintillating bubble chamber, will be competitive to many diverse New Physics scenarios. The scintillating bubble chamber detector could achieve a competitive and comprehensive physics programme with setups in either research reactors or power commercial reactors.

\section{Acknowledgements}
\begin{acknowledgments}
\indent The authors would like to thank the SBC collaboration for useful discussions and for providing details of the Scintillating Bubble Chamber detector.\\
\indent This work is supported by the German-Mexican research collaboration grant SP 778/4-1 (DFG) and 278017 (CONACYT), the projects CONACYT CB-2017-2018/A1-S-13051 and CB-2017-2018/A1-S-8960,  DGAPA UNAM grants PAPIIT-IN107621, PAPIIT-IN107118 and PAPIIT-IN108020, and Fundación Marcos Moshinsky.\\

\end{acknowledgments}
\bibliographystyle{apsrev4-1}
\bibliography{coherentLHC}
\end{document}